\documentclass[11pt]{article}
\usepackage[top=1in, bottom=1in, left=1in, right=1in]{geometry}
\usepackage{amssymb, amsmath, amsthm}  
\usepackage{mathrsfs, amsfonts, bbm}        
\usepackage{graphicx, subfigure}  
\usepackage{enumerate}  
\usepackage{extarrows}            
\usepackage{multirow}             
\usepackage{fullpage}
\usepackage{setspace}
\usepackage{xcolor}
\usepackage{caption}
\usepackage{float}
\usepackage{verbatim}             
\usepackage{soul}
\usepackage{amsmath}
\usepackage{titling}
\usepackage{comment}
\usepackage{booktabs}
\usepackage{multicol}
\allowdisplaybreaks[4]

\usepackage{natbib} 

\usepackage{hyperref} 
\definecolor{darkblue}{rgb}{0,0,.4}
\hypersetup{
colorlinks=true,
citecolor=blue,
urlcolor=blue,
linkcolor=blue,
}

\doublespace

\newcommand{\oline}{\overline}
\newcommand{\mbf}{\mathbf}

\newcommand{\bb}{\mathbb}
\newcommand{\bbm}{\mathbbm}

\newcommand{\tq}{\triangleq}

\DeclareMathAlphabet{\mathpzc}{OT1}{pzc}{m}{it}

\soulregister\cite7 
\soulregister\citep7 
\soulregister\citet7 
\soulregister\ref7 
\soulregister\pageref7 

\begin{document}
\title{Recursive Score and Hessian Computation in Regime-Switching Models}

\author{Chaojun Li\thanks{Academy of Statistics and Interdisciplinary Sciences, Faculty of Economics and Management, 3663 North Zhongshan Road, Shanghai 200062, China. Email: cjli@fem.ecnu.edu.cn}\\{ East China Normal University} \and Shi Qiu \thanks{Depart of Economics, American University of Sharjah. E-mail: shiqiuecon@gamil.com.} \\{American University of Sharjah}}

\date{}
\maketitle

\begin{abstract}
    This study proposes a recursive and easy-to-implement algorithm to compute the score and Hessian matrix in general regime-switching models. We use simulation to compare the asymptotic variance estimates constructed from the Hessian matrix and the outer product of the score. The results favor the latter. 
\end{abstract}



\newpage

\section{Introduction}
Regime-switching models have been widely applied since \citet{hamilton1989new} to analyze how time-series patterns shift across different latent economic states. These models are typically estimated via maximum likelihood, with the computation of the score vector and the Hessian matrix playing a central role in producing reliable standard errors and enabling hypothesis tests. Although \citet{hamilton1996specification} provides an algorithm for computing the score using smoothed probabilities, a direct and generally accepted method for calculating the Hessian matrix remains lacking in the literature. At the same time, inference based on the Hessian matrix is common practice. 
Consequently, applied research often resorts to numerical approximations for inference, as seen in studies by \citet{psaradakis1998finite} and \citet{pouzo2022maximum}.

The algorithm for smoothed functionals in \citet[chapter 4.1]{cappe2005inference} can, in principle, be used to compute both the score and the Hessian. However, it remains unfamiliar to most researchers for several reasons. First, its illustration has been limited to a simple one-parameter model, and although generalizable, the complete algorithm has not been laid out in a general setting. Implementing it requires a preliminary step of applying Fisher’s and Louis’ identities (\citealp[proposition 10.1.6]{cappe2005inference}) to express the score as the conditional expectation of the complete score and to relate the Hessian to that of the complete-data log-likelihood. These expressions must then be reformulated into the specific form required by the smoothing functional algorithm (\citealp[proposition 4.1.3]{cappe2005inference}). For a score vector of dimension $k\times 1$
and a Hessian of size $k\times k$, researchers must derive smoothing functionals for each element, a process that can become cumbersome and tedious.

The contributions of this study are twofold. First, we propose a unified recursive framework to compute the score and Hessian matrix simultaneously. This framework is designed for a general regime-switching specification that encompasses many useful models in the empirical literature as special cases. Unlike the approach in \citet{hamilton1996specification}, our algorithm does not require computing or storing smoothed probabilities and can be integrated directly into the prediction and update steps of likelihood evaluation. It is straightforward to implement, as it avoids the need for preliminary transformations via Fisher’s or Louis’ identities and works directly with derivatives of the period likelihood function. Researchers need only supply the first and second derivatives of certain quantities, and our algorithm returns the score and Hessian, significantly reducing pre-computation effort compared to the smoothed functional approach.
Second, to our knowledge, no thorough investigation has been conducted to compare inference based on the exact score versus the exact Hessian in this context. We supplement our methodological contribution with simulation evidence to examine their finite-sample properties.

{\bf Notations: }let  $\mbf{Y}_m^n\tq(Y_{m},\ldots,Y_n)'$ for $m\leq n$, and let $\mbf{Y}_m^n$ be an empty set for $m>n$. Similarly, define $\mbf{S}_m^n$ and $\mbf{X}_m^n$. We use $\mathbb{N}(\mu,\sigma^2)$ to denote the density of the normal distribution with mean $\mu$ and variance $\sigma^2$.

\section{Model}\label{secmodel}
The conditional distribution of the observable process $Y_t$ is governed by a latent regime process $S_t$, and it admits a density
\begin{align*}
    g_\theta(Y_t|\mbf{S}_{t-p}^t,\mbf{Y}_{t-q}^{t-1},X_t),
\end{align*}
where $p,q\geq0$, $X_t$ is a predetermined variable (vector), $g_\theta$ is a family of density functions indexed by $\theta$.
The regime $S_t$ is a first-order Markov chain on $\bb{S}=\{1,2,\ldots,J\}$ with transition probability 
$q_{ij}=q_\theta(S_t=j|S_{t-1}=i)$ for $i,j=1,\ldots,J$. 
Given $(\mbf{S}_{t-\max\{p,1\}}^{t-1},\mbf{Y}_{t-q}^{t-1},X_t)$, $(Y_{t},S_{t})$ is independent of $(\{Y_k\}_{k\leq t-q-1},\{S_k\}_{k\leq t-\max\{p,q\}},\{X_k\}_{k\leq t-1})$, and the conditional density can be written as
\begin{align}\label{eq:period}
    \mathrm{f}_\theta(Y_t,S_t|\mbf{Y}_{t-q}^{t-1},\mbf{S}_{t-\max\{p,1\}}^{t-1},X_t)=g_\theta(Y_t|\mbf{S}_{t-p}^t,\mbf{Y}_{t-q}^{t-1},X_t)q_\theta(S_t|S_{t-1}).
\end{align}
For short notation, let $\oline{\mbf{S}}_t\tq \mbf{S}_{t-\max\{p,1\}+1}^{t}$. We abbreviate the dependence on $Y_{t}$, $\oline{\mbf{Y}}_{t-1}$, and $X_t$ into the subscript $t$. Then (\ref{eq:period}) can be expressed as $\mathrm{f}_{\theta,t}(S_t,\oline{\mbf{S}}_{t-1})$. Its first and second derivatives are expressed as $\nabla_\theta \mathrm{f}_{\theta,t}(S_t,\oline{\mbf{S}}_{t-1})$ and $\nabla_\theta^2 \mathrm{f}_{\theta,t}(S_t,\oline{\mbf{S}}_{t-1})$.

\noindent {\bf Example.} (Autoregressive model with regime switching) Consider the following model:
\begin{align}
    Y_t -\mu_{s_t} = \phi(Y_{t-1}-\mu_{s_{t-1}})+\sigma_{s_t} u_t,\label{model:musig}
\end{align}
where $u_t$ is an independent and identically distributed (i.i.d.) sequence of random variables. The parameters $\mu_{s_t}$ and $\sigma_{s_t}$ depend on a latent regime variable $s_t$. $s_t=1$ or 2 and follows a first-order Markov chain. 
In this example, $p=q=2$, and
\begin{align*}
    \mathrm{f}_{\theta,t}(S_t,S_{t-1})=\mathrm{f}_{\theta}(Y_t,S_{t}|Y_{t-1},S_{t-1})=\bb{N}(\mu_{s_t}+\phi(Y_{t-1}-\mu_{s_{t-1}}),\sigma_{s_{t}}^2)q_{s_{t-1},s_t}.
\end{align*}
This model was employed in \citet{psaradakis1998finite} to study the finite-sample properties of the maximum likelihood estimator.

This study provides the algorithms to compute the score and hessian matrix of the log-likelihood function given the initial observations $\mbf{Y}^0_{-q+1}$, $\ell_{n,\nu}(\theta)=\log p_{\theta,\nu}(Y_1,\ldots,Y_n|\mbf{Y}^0_{-q+1},\mbf{X}_1^n)$, where $\nu$ is the short notation for the initial distribution of regimes $\nu_\theta(\oline{\mbf{S}}_0)\tq p_\theta(\oline{\mbf{S}}_0|\mbf{Y}_{-q+1}^0)$. Examples of the initial distributions include a fixed starting point $\oline{\mbf{s}}_0$, $\nu(\oline{\mbf{S}}_0)=\bbm{1}\{\oline{\mbf{S}}_0=\oline{\mbf{s}}_0\}$. If the Markov chain is ergodic, the initial distribution can also be chosen to be the unconditional distribution.
The log-likelihood function can be expressed as
\begin{align}
    \ell_{n,\nu}(\theta)
    &=\log \sum_{\oline{\mbf{S}}_0} \sum_{\mbf{S}_1^n} \big(p_{\theta}(Y_1,\ldots,Y_n,S_1,\ldots,S_n|\mbf{Y}^0_{-q+1},\oline{\mbf{S}}_0,\mbf{X}_1^n)\nu_\theta(\oline{\mbf{S}}_0)\big)\nonumber
    \\
    &=\log \sum_{\oline{\mbf{S}}_0}\sum_{\mbf{S}_1^n}
    \big(\prod_{t=1}^n \mathrm{f}_\theta(Y_t,S_t|\mbf{Y}_{t-q}^{t-1},\oline{\mbf{S}}_{t-1},X_t)\nu_\theta(\oline{\mbf{S}}_0)\big).\label{eq:llk}
\end{align}
For notational simplicity, we slightly abuse notation and let $\mathrm{f}_{\theta,0}(S_0,\oline{\mbf{S}}_{-1})$ denote $\nu(\oline{\mbf{S}}_{0})$, where $\oline{\mbf{S}}_{-1}\tq \mbf{S}^{-1}_{-\max\{p,1\}+1}$.
Then (\ref{eq:llk}) can be expressed as
\begin{align*}
    \ell_{n,\nu}(\theta) = \log \sum_{\oline{\mbf{S}}_0}\sum_{\mbf{S}_1^n}
    \big(\prod_{t=0}^n \mathrm{f}_{\theta,t}(S_t,\oline{\mbf{S}}_{t-1})\big).
\end{align*}
We use $\mathrm{p}_{\theta,n}$ as a short notation for the likelihood $p_{\theta,\nu}(Y_1,\ldots,Y_n|\oline{\mbf{Y}}_0,\mbf{X}_1^n)$. Then $\ell_{n,\nu}(\theta)=\log \mathrm{p}_{\theta,n}$.
The score function can be expressed as
\begin{align}
    \nabla_\theta \ell_{n,\nu}(\theta)
    =\frac{\mathrm{s}_{\theta,n}}{\mathrm{p}_{\theta,n}},\label{defscore}
\end{align}
where
\begin{align}
    \mathrm{p}_{\theta,n}
    &=\sum_{\oline{\mbf{S}}_0}\sum_{\mbf{S}_1^n}\big(\prod_{t=0}^n \mathrm{f}_{\theta,t}(S_{t},\oline{\mbf{S}}_{t-1})\big)\label{defp}\\
    \mathrm{s}_{\theta,n}
    &\tq\nabla_\theta \mathrm{p}_{\theta,n}
    =\sum_{\oline{\mbf{S}}_0}\sum_{\mbf{S}_1^n}
    \sum_{t=0}^n\big(\nabla_\theta \mathrm{f}_{\theta,t}(S_t,\oline{\mbf{S}}_{t-1})\times 
    \prod_{0\leq k\leq n,k\neq t} \mathrm{f}_{\theta,k}(S_k,\oline{\mbf{S}}_{k-1})\big).\label{defs}
\end{align}
The Hessian matrix can be expressed as
\begin{align}
    \nabla^2_\theta\ell_{n,\nu}(\theta)
    =\frac{\mathrm{p}_{\theta,n}\nabla_\theta^2\mathrm{p}_{\theta,n}
    -\nabla_\theta \mathrm{p}_{\theta,n}\nabla_\theta^T \mathrm{p}_{\theta,n}}
    {\mathrm{p}_{\theta,n}^2},\label{defhessian}
\end{align}
where $\nabla_\theta^2\mathrm{p}_{\theta,n}=\mathrm{H}_{\theta,n}+\mathrm{h}_{\theta,n}+\mathrm{h}_{\theta,n}^T$ and
\begin{align}
    \mathrm{H}_{\theta,n}
    &\tq\sum_{\oline{\mbf{S}}_0}\sum_{\mbf{S}_1^n}\sum_{t=0}^n\big(\nabla^2_\theta \mathrm{f}_{\theta,t}(S_t,\oline{\mbf{S}}_{t-1})\times 
    \prod_{0\leq k\leq n,k\neq t} \mathrm{f}_{\theta,k}(S_k,\oline{\mbf{S}}_{k-1})\big)\label{defH}\\
    \mathrm{h}_{\theta,n}
    &\tq\sum_{\oline{\mbf{S}}_0}\sum_{\mbf{S}_1^n}\sum_{0\leq t_1<t_2\leq n}\big(\nabla_\theta \mathrm{f}_{\theta,t_1}(S_{t_1},\oline{\mbf{S}}_{t_1-1})\nabla_{\theta}^T\mathrm{f}_{\theta,t_2}(S_{t_2},\oline{\mbf{S}}_{t_2-1})
    \prod_{0\leq k\leq n,k\neq t_1,t_2}\mathrm{f}_{\theta,k}(S_k,\oline{\mbf{S}}_{k-1})\big).\label{defh}
\end{align}
The next section presents the algorithm used to compute (\ref{defscore}) and (\ref{defhessian}). Note that the algorithm requires researchers to supply only (\ref{eq:period}), the initial distribution $\nu_\theta(\oline{\mbf{S}}_t)$, and their first and second derivatives. Unlike the smoothed functional approach of \citet{hamilton1996specification} and \citet[Chapter 4.1]{cappe2005inference}, our method does not require any preliminary transformations using Fisher's or Louis' identities. This significantly reduces the pre-computation effort.

\section{Algorithm to compute score and Hessian}\label{secalgorithm}
We can compute the score and Hessian matrix if we know how to compute (\ref{defp})--(\ref{defs}) and (\ref{defH})--(\ref{defh}). 
(\ref{defH}) can be computed as (\ref{defs}) by replacing $\nabla_\theta \mathrm{f}_{\theta,t}(S_t,\oline{\mbf{S}}_{t-1})$ with $\nabla_\theta^2 \mathrm{f}_{\theta,t}(S_t,\oline{\mbf{S}}_{t-1})$. Thus, the main challenge is to compute (\ref{defp}), (\ref{defs}), and (\ref{defh}) efficiently. 
To explain the idea of the algorithm, we first consider the simple model where there is no regime switching in Section \ref{subsimple}.

\subsection{Simple model without regime switching}\label{subsimple}
In the simple model without regime switching, (\ref{defp}), (\ref{defs}), and (\ref{defh}) are simplified to
\begin{align}
    \mathrm{p}_{\theta,t}
    &=\prod_{k=0}^t\mathrm{f}_{\theta,k}\nonumber\\
    \mathrm{s}_{\theta,t}
    &=\sum_{k=0}^t\big( \mathrm{f}_{\theta,0}\ldots
    \mathrm{f}_{\theta,k-1}
    \times\nabla_\theta \mathrm{f}_{\theta,k}\times \mathrm{f}_{\theta,k+1}\ldots
    \mathrm{f}_{\theta,t})\label{defssimple}\\
    \mathrm{h}_{\theta,t}
    &=\sum_{0\leq t_1<t_2\leq t}\big(\nabla_\theta \mathrm{f}_{\theta,t_1}\nabla_{\theta}^T\mathrm{f}_{\theta,t_2}\prod_{1\leq k\leq t,k\neq t_1,t_2}\mathrm{f}_{\theta,k}\big)\nonumber\\
    &=\mathrm{s}_{\theta,0}\nabla^T_\theta\mathrm{f}_{\theta,1}\mathrm{f}_{\theta,2}\cdots\mathrm{f}_{\theta,t}
    +\mathrm{s}_{\theta,1}\nabla^T_\theta\mathrm{f}_{\theta,2}\mathrm{f}_{\theta,3}\cdots\mathrm{f}_{\theta,t}
    +\ldots
    +\mathrm{s}_{\theta,t-1}\nabla^T_\theta\mathrm{f}_{\theta,t}.
    \label{defhsimple}
\end{align}
They can be computed according to the following algorithm.
\begin{enumerate}
    \item Initialization: for $t=1$
    \begin{align*}
        \mathrm{p}_{\theta,1}=\mathrm{f}_{\theta,0}\mathrm{f}_{\theta,1},\quad
        \mathrm{s}_{\theta,1}=\nabla_\theta \mathrm{f}_{\theta,0}\mathrm{f}_{\theta,1}
        +\mathrm{f}_{\theta,0}\nabla_\theta\mathrm{f}_{\theta,1},\quad
        \mathrm{h}_{\theta,1}=\nabla_\theta \mathrm{f}_{\theta,0}\nabla_\theta^T\mathrm{f}_{\theta,1}.
    \end{align*}
    \item Recursion: for $t=2,\ldots,n$, 
    \begin{align}
        \mathrm{p}_{\theta,t}
        &=\mathrm{p}_{\theta,t-1}\times \mathrm{f}_{\theta,t}\nonumber\\
        \mathrm{s}_{\theta,t}
        &=\mathrm{s}_{\theta,t-1}\times \mathrm{f}_{\theta,t}
        +\mathrm{p}_{\theta,t-1}\times \nabla_\theta \mathrm{f}_{\theta,t}
        \label{supdatesimple}\\
        \mathrm{h}_{\theta,t}
        &=\mathrm{h}_{\theta,t-1}\times \mathrm{f}_{\theta,t}+\mathrm{s}_{\theta,t-1}\nabla_\theta^T\mathrm{f}_{\theta,t}
        \label{hupdatesimple}
    \end{align}
\end{enumerate}
To see why we update as (\ref{supdatesimple}), notice that each summand in (\ref{defssimple}) is the product of $t+1$ terms, with one being the the derivative and the rest being the period likelihood. The summand in $\mathrm{s}_{\theta,t-1}$ already contains one derivative, so we multiply it with the period likelihood function $\mathrm{f}_{\theta,t}$, which brings the first term in (\ref{supdatesimple}). We add the second term in (\ref{supdatesimple}) to include the term with derivative at time $t$. To write it more specifically,
\begin{align*}
    t=2, &&
    \mathrm{s}_{\theta,2}
    &=\mathrm{s}_{\theta,1}\times \mathrm{f}_{\theta,2}
    +\mathrm{p}_{\theta,1}\times \nabla \mathrm{f}_{\theta,2}
    =(\nabla_{\theta}\mathrm{f}_{\theta,0}\mathrm{f}_{\theta,1}
    +\mathrm{f}_{\theta,0}\nabla_{\theta}\mathrm{f}_{\theta,1})\mathrm{f}_{\theta,2}+\mathrm{f}_{\theta,0}\mathrm{f}_{\theta,1}\nabla_\theta\mathrm{f}_{\theta,2}\\
    \vdots &&   &\vdots \\
    t=n, &&
    \mathrm{s}_{\theta,n}
    &=\mathrm{s}_{\theta,n-1}\times \mathrm{f}_{\theta,n}
    +\mathrm{p}_{\theta,n-1}\times \nabla \mathrm{f}_{\theta,n}\\
    &&
    &=(\nabla_{\theta}\mathrm{f}_{\theta,0}\mathrm{f}_{\theta,1}\ldots\mathrm{f}_{\theta,n-1}
    +\ldots
    +\mathrm{f}_{\theta,0}\cdots\mathrm{f}_{\theta,n-2}\nabla_{\theta}\mathrm{f}_{\theta,n-1})\mathrm{f}_{\theta,n}+\mathrm{f}_{\theta,0}\cdots\mathrm{f}_{\theta,n-1}\nabla_\theta\mathrm{f}_{\theta,n}\\
    &&
    &=\nabla_\theta \mathrm{f}_{\theta,0}\mathrm{f}_{\theta,1}\cdots\mathrm{f}_{\theta,n}
    +\mathrm{f}_{\theta,0}\nabla_\theta \mathrm{f}_{\theta,1}\mathrm{f}_{\theta,2} \cdots\mathrm{f}_{\theta,n}
    +\ldots
    +\mathrm{f}_{\theta,0}\cdots \mathrm{f}_{\theta,n-1}\nabla_\theta \mathrm{f}_{\theta,n}.
\end{align*}
(\ref{hupdatesimple}) is updated with the same reasoning. Each summand in (\ref{defhsimple}) is the product of $t+1$ terms, with two being the derivatives and the rest being the period likelihood. The summand in $\mathrm{h}_{\theta,t-1}$ already contains two derivatives, so we multiply it with $\mathrm{f}_{\theta,t}$, which is the first term in (\ref{hupdatesimple}). For the second term in (\ref{hupdatesimple}), the summand in $\mathrm{s}_{\theta,t-1}$ contains one derivative, so we multiply it with $\nabla_\theta^T\mathrm{f}_{\theta,t}$. Next subsection explains the algorithm to compute the score and Hessian matrix for regime-switching models.

\subsection{The algorithm to compute the score and Hessian matrix}\label{subalgorithm1}
Section \ref{subsimple} explains the algorithm to compute (\ref{defp}), (\ref{defs}), and (\ref{defh}) in the model without regime switching.
To obtain the complete algorithm for the model with regime switching, we need to make two modifications. First, for models with regime switching, we need to take summation over regimes at appropriate steps. 
For each regime $S_t$, 
it is involved in $\mathrm{f}_{\theta,k}(S_{k},\oline{\mbf{S}}_{k-1})$, $t\leq k\leq t+p$, but not in $\mathrm{f}_{\theta,\ell}(S_{\ell},\oline{\mbf{S}}_{\ell-1})$, $\ell\geq t+p+1$. Thus, we can take summation over $S_t$ at the recursion step of $t+p$. Second, in (\ref{defscore}) and (\ref{defhessian}), the computation of the denominator $\mathrm{p}_{\theta,n}=p_{\theta,\nu}(Y_1,\ldots,Y_n|\oline{\mbf{Y}}_0,\mbf{X}_1^n)$ might suffer from numerical overflow or underflow.\footnote{To see this more clearly, note that under suitable regularity conditions, the normalized log-likelihood satisfies
$n^{-1}\log p_{\theta,\nu}(Y_1,\ldots,Y_n\mid\overline{\mathbf{Y}}_0,\mathbf{X}_1^n)\overset{p}{\longrightarrow}\ell(\theta)$
in the consistency proof; see \citet{douc2004asymptotic} and \citet{kasahara2019asymptotic}. Consequently, for large $n$, the likelihood itself behaves as
$p_{\theta,\nu}(Y_1,\ldots,Y_n\mid\overline{\mathbf{Y}}_0,\mathbf{X}_1^n)\approx\exp\big(n\ell(\theta)\big)$.
If $\ell(\theta)<0$, then $\exp(n\ell(\theta))$ decays exponentially to zero. Once its magnitude falls below the machine epsilon, floating‑point software will round it to zero, resulting in numerical underflow. Conversely, if $\ell(\theta)>0$, the term grows exponentially and may exceed the largest representable floating‑point number, causing numerical overflow (stored as INF).} Thus, instead of making a division in the last step, we re-scale each value with the period likelihood at each recursion step. 
From $\mathrm{p}_{\theta,t}=\prod_{k=1}^tp_{\theta,\nu}(Y_k|\oline{\mbf{Y}}_0^{k-1},\mbf{X}_1^k)$, we can re-scale at each recursion with $p_{\theta,\nu}(Y_k|\oline{\mbf{Y}}_0^{k-1},\mbf{X}_1^k)$.
The complete algorithm can be described as follows. 
\begin{enumerate}
    \item Initialization: for $t=1$,
    \begin{align*}
        \mathpzc{p}_{\theta,1}(\oline{\mbf{S}}_1)
        &=\sum_{S_{-\max\{p,1\}+1}}\mathrm{f}_{\theta,0}(S_0,\oline{\mbf{S}}_{-1})
        \mathrm{f}_{\theta,1}(S_1,\oline{\mbf{S}}_0)\\
        \mathpzc{s}_{\theta,1}(\oline{\mbf{S}}_1)
        &=\sum_{S_{-\max\{p,1\}+1}}\nabla_\theta \mathrm{f}_{\theta,0}(S_0,\oline{\mbf{S}}_{-1})\mathrm{f}_{\theta,1}(S_1,\oline{\mbf{S}}_{0})
        +\mathrm{f}_{\theta,0}(S_0,\oline{\mbf{S}}_{-1})\nabla_\theta \mathrm{f}_{\theta,1}(S_1,\oline{\mbf{S}}_0)\\
        \mathpzc{h}_{\theta,1}(\oline{\mbf{S}}_1)
        &=\sum_{S_{-\max\{p,1\}+1}}\nabla_\theta\mathrm{f}_{\theta,0}(S_0,\oline{\mbf{S}}_{-1})\nabla_\theta^T\mathrm{f}_{\theta,1}(S_1,\oline{\mbf{S}}_0)\\
        \mathpzc{H}_{\theta,1}(\oline{\mbf{S}}_1)
        &=\sum_{S_{-\max\{p,1\}+1}}\nabla_\theta^2 \mathrm{f}_{\theta,0}(S_0,\oline{\mbf{S}}_{-1})\mathrm{f}_{\theta,1}(S_1,\oline{\mbf{S}}_0)
        +\mathrm{f}_{\theta,0}(S_0,\oline{\mbf{S}}_{-1})\nabla_\theta^2 \mathrm{f}_{\theta,1}(S_1,\oline{\mbf{S}}_0)
    \end{align*}
    Re-scale:
    $\widetilde{\mathpzc{p}}_{\theta,1}(\oline{\mbf{S}}_1)
    =\frac{\mathpzc{p}_{\theta,1}(\oline{\mbf{S}}_1)}{\sum_{\oline{\mbf{S}}_1}\mathpzc{p}_{\theta,1}(\oline{\mbf{S}}_1)}$,
    $\widetilde{\mathpzc{s}}_{\theta,1}(\oline{\mbf{S}}_1)
    =\frac{\mathpzc{s}_{\theta,1}(\oline{\mbf{S}}_1)}{\sum_{\oline{\mbf{S}}_1}\mathpzc{p}_{\theta,1}(\oline{\mbf{S}}_1)}$,
    $\widetilde{\mathpzc{h}}_{\theta,1}(\oline{\mbf{S}}_1)
    =\frac{\mathpzc{h}_{\theta,1}(\oline{\mbf{S}}_1)}{\sum_{\oline{\mbf{S}}_1}\mathpzc{p}_{\theta,1}(\oline{\mbf{S}}_1)}$,
    $\widetilde{\mathpzc{H}}_{\theta,1}(\oline{\mbf{S}}_1)=\frac{\mathpzc{H}_{\theta,1}(\oline{\mbf{S}}_1)}{\sum_{\oline{\mbf{S}}_1}\mathpzc{p}_{\theta,1}(\oline{\mbf{S}}_1)}$.
    \item Recursion: for $t=2,\ldots,n$, 
    \begin{align*}
        \mathpzc{H}_{\theta,t}(\oline{\mbf{S}}_t)
        &=\sum_{S_{t-p}}\big[\widetilde{\mathpzc{H}}_{\theta,t-1}(\oline{\mbf{S}}_{t-1})\times \mathrm{f}_{\theta,t}(S_t,\oline{\mbf{S}}_{t-1})
        +\widetilde{\mathpzc{p}}_{\theta,t-1}(\oline{\mbf{S}}_{t-1})\times \nabla_\theta^2 \mathrm{f}_{\theta,t}(S_t,\oline{\mbf{S}}_{t-1})\big]\\
        \mathpzc{h}_{\theta,t}(\oline{\mbf{S}}_t)
        &=\sum_{S_{t-p}}\big[\widetilde{\mathpzc{h}}_{\theta,t-1}(\oline{\mbf{S}}_{t-1})\times \mathrm{f}_{\theta,t}(S_t,\oline{\mbf{S}}_{t-1})
        +\widetilde{\mathpzc{s}}_{\theta,t-1}(\oline{\mbf{S}}_{t-1})\times \nabla_\theta^T \mathrm{f}_{\theta,t}(S_t,\oline{\mbf{S}}_{t-1})\big]\\
        \mathpzc{s}_{\theta,t}(\oline{\mbf{S}}_t)
        &=\sum_{S_{t-p}}\big[\widetilde{\mathpzc{s}}_{\theta,t-1}(\oline{\mbf{S}}_{t-1})\times \mathrm{f}_{\theta,t}(S_t,\oline{\mbf{S}}_{t-1})
        +\widetilde{\mathpzc{p}}_{\theta,t-1}(\oline{\mbf{S}}_{t-1})\times \nabla_\theta \mathrm{f}_{\theta,t}(S_t,\oline{\mbf{S}}_{t-1})\big]\\
        \mathpzc{p}_{\theta,t}(\oline{\mbf{S}}_t)
        &=\sum_{S_{t-p}}\widetilde{\mathpzc{p}}_{\theta,t-1}(\oline{\mbf{S}}_{t-1})\times \mathrm{f}_{\theta,t}(S_t,\oline{\mbf{S}}_{t-1})
    \end{align*}
    Re-scale:
    $\widetilde{\mathpzc{p}}_{\theta,t}(\oline{\mbf{S}}_{t})=\frac{\mathpzc{p}_{\theta,t}(\oline{\mbf{S}}_{t})}{\sum_{\oline{\mbf{S}}_{t}}\mathpzc{p}_{\theta,t}(\oline{\mbf{S}}_{t})}$,
    $\widetilde{\mathpzc{s}}_{\theta,t}(\oline{\mbf{S}}_{t})=\frac{\mathpzc{s}_{\theta,t}(\oline{\mbf{S}}_{t})}{\sum_{\oline{\mbf{S}}_{t}}\mathpzc{p}_{\theta,t}(\oline{\mbf{S}}_{t})}$,
    $\widetilde{\mathpzc{h}}_{\theta,t}(\oline{\mbf{S}}_{t})=\frac{\mathpzc{h}_{\theta,t}(\oline{\mbf{S}}_{t})}{\sum_{\oline{\mbf{S}}_{t}}\mathpzc{p}_{\theta,t}(\oline{\mbf{S}}_{t})}$,
    $\widetilde{\mathpzc{H}}_{\theta,t}(\oline{\mbf{S}}_{t})=\frac{\mathpzc{H}_{\theta,t}(\oline{\mbf{S}}_{t})}{\sum_{\oline{\mbf{S}}_{t}}\mathpzc{p}_{\theta,t}(\oline{\mbf{S}}_{t})}$.
    \item Let $\mathrm{H}_{\theta,n}=\sum_{\oline{\mbf{S}}_{n}}\widetilde{\mathpzc{H}}_{\theta,n}(\oline{\mbf{S}}_{n})$,
    $\mathrm{h}_{\theta,n}=\sum_{\oline{\mbf{S}}_{n}}\widetilde{\mathpzc{h}}_{\theta,n}(\oline{\mbf{S}}_{n})$,
    $\mathrm{s}_{\theta,n}=\sum_{\oline{\mbf{S}}_{n}}\widetilde{\mathpzc{s}}_{\theta,n}(\oline{\mbf{S}}_{n})$, and
    $\mathrm{p}_{\theta,n}=\sum_{\oline{\mbf{S}}_{n}}\widetilde{\mathpzc{p}}_{\theta,n}(\oline{\mbf{S}}_{n})$.
    Compute the score and the Hessian matrix:
    \begin{align*}
        \nabla_\theta\ell_{n,\nu}(\theta)
        &=\mathrm{s}_{\theta,n}\\
        \nabla_\theta^2\ell_{n,\nu}(\theta)
        &=\mathrm{H}_{\theta,n}+\mathrm{h}_{\theta,n}+\mathrm{h}_{\theta,n}^T-\mathrm{s}_{\theta,n}\mathrm{s}_{\theta,n}^T.
    \end{align*}
\end{enumerate}
Unlike the smoothed functional approach in \citet{hamilton1996specification}, our algorithm is recursive and does not require the computation and storage of the smoothed probabilities.
The connection between our algorithm and the standard prediction-and-update steps for evaluating the likelihood function can be seen as follows. For $t\geq 2$, $\mathpzc{p}_{\theta,t}(\oline{\mbf{S}}_{t})
=p_{\theta,\nu}(Y_t,\oline{\mbf{S}}_{t}|\oline{\mbf{Y}}_0^{t-1},\mbf{X}_1^t)$ 
$\sum_{\oline{\mbf{S}}_{t}}\mathpzc{p}_{\theta,t}(\oline{\mbf{S}}_{t})
=p_{\theta,\nu}(Y_t|\oline{\mbf{Y}}_0^{t-1},\mbf{X}_1^t)$ are the terms computed in the prediction step.
$\widetilde{\mathpzc{p}}_{\theta,t}(\oline{\mbf{S}}_{t})=p_{\theta,\nu}(\oline{\mbf{S}}_{t}|\oline{\mbf{Y}}_0^{t},\mbf{X}_1^t)$ is the term computed in the updating step.
Thus, the recursion in this algorithm extends the standard prediction and update algorithm of likelihood evaluation to compute the score and the Hessian matrix simultaneously. 

\section{Simulation}
\citet{kasahara2019asymptotic} established the asymptotic normality of the MLE in regime-switching models and showed that the asymptotic covariance matrix of $\sqrt{n}(\hat{\theta}_{n,\nu}-\theta_0)$ can be consistently estimated using either the Hessian-based estimator
\begin{align}
    \left(-n^{-1}\nabla_\theta^2\ell_{n,\nu}(\hat{\theta})\right)^{-1}\label{eq:hessian}
\end{align}
or the outer product of the score 
\begin{align}
    \left[n^{-1}\sum_{t=1}^n \left(\nabla_\theta \log p_{\theta,\nu}(Y_t|\mbf{Y}_1^t,\oline{\mbf{Y}}_0,\mbf{X}_1^n)\nabla_\theta^T \log p_{\theta,\nu}(Y_t|\mbf{Y}_1^t,\oline{\mbf{Y}}_0,\mbf{X}_1^n)\right)\right]^{-1}.\label{eq:ops}
\end{align}
This section compares these two methods for estimating the asymptotic covariance matrix. We consider a comprehensive list of autoregressive models with regime switching.
The data-generating process is the fully regime-switching autoregression
\begin{align}
    Y_t -\mu_{s_t} = \phi_{s_t}(Y_{t-1}-\mu_{s_{t-1}})+\sigma_{s_t} u_t,\quad \bb{S}=\{1,2\}\label{model:all}
\end{align}
and its seven nested variants obtained by restricting any subset of $(\mu,\phi,\sigma)$ remain regime-invariant. The resulting models are denoted as $M_{\mu\phi\sigma}$, $M_{\phi\sigma}$, $M_{\mu\sigma}$, $M_{\mu\phi}$, $M_{\mu}$, $M_{\phi}$ and $M_{\sigma}$, where the subscripts indicate which parameters switch. For instance, the model $M_{\mu\phi\sigma}$ is (\ref{model:all}). The model $M_{\mu\sigma}$ is (\ref{model:musig}). \citet{psaradakis1998finite} examined the performance of the inference based on the numerical Hessian matrix for model $M_{\mu\sigma}$. According to their simulation results, the asymptotic approximation is worse if $\phi,q_{11},$ and $q_{22}$ are close to unity. Thus, we consider the following parameters when the asymptotic distribution is not favored. $q_{11}=q_{22}=0.95$. For the parameters that are allowed to switch,
\begin{align*}
    (\mu_1,\mu_2)=(1,5),\qquad (\sigma_1^2,\sigma_2^2)=(1,3),\qquad
    (\phi_1,\phi_2)=(0.2,0.9)
\end{align*}
when the parameters do not switch,
\begin{align*}
    \mu=1,\qquad\sigma^2=1,\qquad\phi=0.9
\end{align*}
For each model, we generate 1000 data sets of sample sizes $n = $100, 200, 400, and 800.\footnote{We simulate $(800 + n)$ periods and use the last $n$ observations as our sample, so that the initial value of the simulated data set is approximately drawn from the
stationary distribution.} For each data set, we compute the MLE using the Broyden-Fletcher-Goldfarb-Shanno optimization method and compute the standard errors based on (\ref{eq:hessian}) and (\ref{eq:ops}). 

\begin{table}[]
\centering
\caption{Coverage frequencies of the 95\% confidence intervals}
\label{tab:simci}
\begin{tabular}{ccccccccccc}
\hline
\multirow{2}{*}{Model} & \multicolumn{5}{c}{Inference based on (\ref{eq:hessian})} & \multicolumn{5}{c}{Inference based on (\ref{eq:ops})} \\ \cmidrule(lr){2-6} \cmidrule(lr){7-11} 
 & \multicolumn{1}{c}{$n$} & 100 & 200 & 400 & 800 & \multicolumn{1}{c}{$n$} & 100 & 200 & 400 & 800 \\ \hline
\multicolumn{1}{c}{\multirow{6}{*}{$M_{\mu}$}} & \multicolumn{1}{c}{$\mu_1$} & \multicolumn{1}{l}{0.804} & \multicolumn{1}{l}{0.866} & \multicolumn{1}{l}{0.914} & \multicolumn{1}{l}{0.925} & \multicolumn{1}{c}{$\mu_1$} & \multicolumn{1}{l}{0.850} & \multicolumn{1}{l}{0.886} & \multicolumn{1}{l}{0.918} & \multicolumn{1}{l}{0.928} \\
\multicolumn{1}{l}{} & \multicolumn{1}{c}{$\mu_2$} & \multicolumn{1}{l}{0.806} & \multicolumn{1}{l}{0.864} & \multicolumn{1}{l}{0.915} & \multicolumn{1}{l}{0.928} & \multicolumn{1}{c}{$\mu_2$} & \multicolumn{1}{l}{0.869} & \multicolumn{1}{l}{0.890} & \multicolumn{1}{l}{0.923} & \multicolumn{1}{l}{0.935} \\
\multicolumn{1}{l}{} & \multicolumn{1}{c}{$\sigma^2$} & \multicolumn{1}{c}{0.891} & \multicolumn{1}{l}{0.918} & \multicolumn{1}{c}{0.936} & \multicolumn{1}{l}{0.941} & \multicolumn{1}{c}{$\sigma^2$} & \multicolumn{1}{c}{0.917} & \multicolumn{1}{l}{0.924} & \multicolumn{1}{c}{0.939} & \multicolumn{1}{c}{0.941} \\
\multicolumn{1}{l}{} & \multicolumn{1}{c}{$\phi$} & \multicolumn{1}{c}{0.860} & \multicolumn{1}{l}{0.915} & \multicolumn{1}{l}{0.917} & \multicolumn{1}{c}{0.942} & \multicolumn{1}{c}{$\phi$} & \multicolumn{1}{l}{0.885} & \multicolumn{1}{l}{0.913} & \multicolumn{1}{l}{0.931} & \multicolumn{1}{c}{0.941} \\
\multicolumn{1}{l}{} & \multicolumn{1}{c}{$q_{11}$} & \multicolumn{1}{c}{0.908} & \multicolumn{1}{l}{0.935} & \multicolumn{1}{l}{0.945} & \multicolumn{1}{c}{0.939} & \multicolumn{1}{c}{$q_{11}$} & \multicolumn{1}{l}{0.963} & \multicolumn{1}{l}{0.945} & \multicolumn{1}{l}{0.947} & \multicolumn{1}{c}{0.944} \\
\multicolumn{1}{l}{} & \multicolumn{1}{c}{$q_{22}$} & \multicolumn{1}{c}{0.937} & \multicolumn{1}{l}{0.933} & \multicolumn{1}{l}{0.949} & \multicolumn{1}{c}{0.955} & \multicolumn{1}{c}{$q_{22}$} & \multicolumn{1}{l}{0.972} & \multicolumn{1}{l}{0.951} & \multicolumn{1}{l}{0.951} & \multicolumn{1}{c}{0.956} \\ \hline
\multirow{6}{*}{$M_{\sigma}$} & \multicolumn{1}{c}{$\mu$} & 0.661 & 0.824 & 0.894 & 0.921 & \multicolumn{1}{c}{$\mu$} & 0.830 & 0.884 & 0.904 & 0.923 \\
 & \multicolumn{1}{c}{$\sigma_1^2$} & 0.567 & 0.759 & 0.865 & 0.917 & \multicolumn{1}{c}{$\sigma_1^2$} & 0.831 & 0.859 & 0.890 & 0.914 \\
 & \multicolumn{1}{c}{$\sigma_2^2$} & 0.699 & 0.871 & 0.928 & 0.950 & \multicolumn{1}{c}{$\sigma_2^2$} & 0.906 & 0.941 & 0.941 & 0.952 \\
 & \multicolumn{1}{c}{$\phi$} & 0.705 & 0.853 & 0.918 & 0.936 & \multicolumn{1}{c}{$\phi$} & 0.899 & 0.916 & 0.929 & 0.939 \\
 & \multicolumn{1}{c}{$q_{11}$} & 0.759 & 0.843 & 0.894 & 0.923 & \multicolumn{1}{c}{$q_{11}$} & 0.964 & 0.914 & 0.907 & 0.918 \\
 & \multicolumn{1}{c}{$q_{22}$} & 0.786 & 0.866 & 0.905 & 0.937 & \multicolumn{1}{c}{$q_{22}$} & 0.938 & 0.911 & 0.914 & 0.935 \\ \hline
\end{tabular}
\end{table}

\begin{table}[]
\centering
\caption{Ratios of sampling standard deviations to estimated standard errors}
\label{tab:simse}
\begin{tabular}{ccccccccccc}
\hline
\multirow{2}{*}{Model} & \multicolumn{5}{c}{Inference based on (\ref{eq:hessian})} & \multicolumn{5}{c}{Inference based on (\ref{eq:ops})} \\ \cmidrule(lr){2-6} \cmidrule(lr){7-11}
 & $n$ & 100 & 200 & 400 & 800 & $n$ & 100 & 200 & 400 & 800 \\ \hline
\multirow{6}{*}{$M_{\mu}$} & $\mu_1$ & 1.492 & 1.254 & 1.137 & 1.050 & $\mu_1$ & 1.293 & 1.175 & 1.082 & 1.030 \\
 & $\mu_2$ & 1.526 & 1.257 & 1.127 & 1.070 & $\mu_2$ & 1.294 & 1.159 & 1.083 & 1.045 \\
 & $\sigma^2$ & 1.127 & 1.092 & 1.062 & 1.013 & $\sigma^2$ & 1.026 & 1.045 & 1.047 & 1.010 \\
 & $\phi$ & 1.395 & 1.207 & 1.132 & 1.043 & $\phi$ & 1.267 & 1.133 & 1.079 & 1.012 \\
 & $q_{11}$ & 3.001 & 1.374 & 1.178 & 1.083 & $q_{11}$ & 2.548 & 1.266 & 1.149 & 1.063 \\
 & $q_{22}$ & 2.750 & 1.616 & 1.200 & 1.056 & $q_{22}$ & 2.254 & 1.462 & 1.149 & 1.029 \\ \hline
\multirow{6}{*}{$M_{\sigma}$} & $\mu$ & 1.700 & 1.190 & 1.137 & 1.081 & $\mu$ & 1.655 & 1.183 & 1.132 & 1.076 \\
 & $\sigma_1^2$ & 2.158 & 1.756 & 1.409 & 1.208 & $\sigma_1^2$ & 1.568 & 1.509 & 1.336 & 1.184 \\
 & $\sigma_2^2$ & 1.548 & 1.459 & 1.271 & 1.117 & $\sigma_2^2$ & 1.121 & 1.184 & 1.144 & 1.070 \\
 & $\phi$ & 1.197 & 1.102 & 1.093 & 1.011 & $\phi$ & 1.149 & 1.081 & 1.083 & 1.008 \\
 & $q_{11}$ & 2.467 & 3.312 & 2.714 & 1.578 & $q_{11}$ & 1.831 & 2.929 & 2.540 & 1.488 \\
 & $q_{22}$ & 2.328 & 3.189 & 1.952 & 1.715 & $q_{22}$ & 2.016 & 2.714 & 1.809 & 1.645 \\ \hline
\end{tabular}
\end{table}

We report the coverage frequencies of the 95\% confidence intervals in Table \ref{tab:simci} and the ratio of the standard deviation of the MLE to the median of the standard errors across 1000 datasets in Table \ref{tab:simse} for $M_\mu$ and $M_\sigma$. 
The asymptotic normality provides a better approximation in $M_\mu$ than $M_{\sigma}$. More specifically, the coverage frequencies are closer to the nominal level of 95\% and the ratios of sampling standard deviations to the estimated standard errors are closer to unity in the model $M_\mu$ than $M_\sigma$. The performance in the remaining models falls between these two cases. To save space, we present only results for $M_\mu$ and $M_\sigma$. Full results are available upon request.
When $n=100$, the performance of the asymptotic normal distribution is somewhat discouraging. The coverage frequencies are lower than the nominal level of 95\%, and the estimated standard errors are smaller than the sampling standard deviations of the MLE. When $n$ gets larger to 800, the performance improves. Moreover, inference based on the outer product of the score (\ref{eq:ops}) outperforms that based on (\ref{eq:hessian}), in the sense that coverage frequencies are closer to 95\% and the ratio of standard deviations to standard errors is closer to unity. This pattern holds across all models examined.

\section{Conclusion}

This study proposes a recursive algorithm for computing the score vector and Hessian matrix in regime-switching models, without relying on pre-computation via Fisher’s or Louis’ identities. Simulation results show that asymptotic approximations based on the outer product of the score are favored over those derived from the Hessian matrix. While the presented framework is general, it does not encompass more complex specifications such as regime-switching GARCH models or regime-switching state-space models, where likelihood evaluation requires approximation. Extending the algorithm to compute score and Hessian matrix for such models will be addressed in future work.

\bibliographystyle{asa}
\bibliography{Reference_ArXiv}

\end{document}